\documentclass[12pt]{article}

\usepackage{bm}
\usepackage{amsmath}
\usepackage{amssymb}

\usepackage{graphicx}
\usepackage{float}
\usepackage{slashbox}
\usepackage{color}
\usepackage{multirow}

\renewcommand{\a}{{\mathbf a}}
\renewcommand{\b}{{\mathbf b}}
\newcommand{\cvec}{{\mathbf c}}
\renewcommand{\d}{{\mathbf d}}

\renewcommand{\j}{{\mathbf j}}

  \newcommand{\m}{{\mathbf m}}

\renewcommand{\r}{{\mathbf r}}

  \newcommand{\E}{{\mathbf E}}
  \newcommand{\B}{{\mathbf B}}
  \newcommand{\A}{{\mathbf A}}
  \newcommand{\C}{{\mathbf C}}
  
  \newcommand{\F}{{\mathbf F}}
  \newcommand{\T}{{\mathbf T}}
\renewcommand{\phi}{{\varphi}}

\def\div{\mathop{\rm div}\nolimits}
\def\rot{\mathop{\rm rot}\nolimits}
\def\bnabla{{\bm\nabla}}

\newcommand*{\cdt}[1]{\overset{\mbox{\tiny$\bullet$}}{#1}}
\newcommand*{\cddtt}[1]{\overset{\mbox{\tiny$\bullet\bullet$}}{#1}}

\begin{document}

\title{Multipole expansions for time-dependent charge and current distributions in quasistatic approximation}

\author{Yuri Krynytskyi and Andrij Rovenchak\\
Department for Theoretical Physics,\\
Ivan Franko National University of Lviv,\\
12, Drahomanov Street, Lviv, 79005, Ukraine,\\
e-mail: \texttt{\{yurikryn,andrij.rovenchak\}@gmail.com}}
\maketitle

\begin{abstract}
We propose a consistent approach to the definition of electric, magnetic, and toroidal multipole moments. Electric and magnetic fields are split into potential, vortex, and radiative terms, with the latter ones dropped off in the quasistatic approximation. The potential part of the electric field, the vortex parts of the magnetic field and vector potential contain gradients of scalar functions. Formally introducing magnetic and toroidal analogs of the electric charge, we apply multipole expansions  for those scalars. Closed-form expressions are derived in an arbitrary order for electric, magnetic, and toroidal multipoles, which constitute a full system for expansions of the electromagnetic field.

\textbf{Key words:} {multipole expansion; higher multipole moments; toroidal multipoles; quasistatic approximation.}

\medskip
PACS number: 03.50.De
\end{abstract}

\section{Introduction}

Multipole expansions are efficient techniques for calculations of electromagnetic fields generated by complex distributions of charges and currents \cite{Griffiths:1999,Jackson:1999}. In problems of electro- and magnetostatics, expansions of scalar and vector potentials generating electric and magnetic multipoles are sufficient. Time-dependent distributions, however, generate a third type of multipoles generally known as toroidal moments. The latter appeared recently, in particular, in the studies from the domain of nanophotonics and metamaterials, see \cite{Li&Crozier:2018} and references therein.
A vast variety of applications involving toroidal structures was analyzed in another review\cite{Parasimakis_etal:2016}, from toroidal moments in  atomic nuclei or biological objects like archeon and red blood cells, to protection of superconducting qubits from external influences.

The approaches to the analysis of toroidal contributions vary in the literature \cite{Dubovik&Tosunyan:1983,Radescu&Vaman:2002,Vrejoiu&Zus:2010,Ross:2012,Nanz:2016,Alaee_etal:2018}.
Following \cite{Dubovik&Cheshkov:1974}, the authors of \cite{Radescu&Vaman:2002} provide the exact multipole expansions for the radiation intensity, angular momentum loss, and the recoil force via electric, magnetic, and toroid mean square radii. The authors of \cite{Vrejoiu&Zus:2009} start from the Jefimenko's equations and thus present multipole expansions of electric and magnetic fields rather than traditional analysis of scalar and vector potentials. Multipole expansion of the action was applied to obtain expressions for the electric and magnetic multipole moments in \cite{Ross:2012}. Exact formulas for multipoles of an arbitrary order, including toroidal multipoles, were obtained in \cite{Radescu&Vaman:2012}. In recent papers \cite{Fernandez-Corbaton_etal:2017,Alaee_etal:2018}, the terms corresponding to the toroidal moments were included in the definitions of the electric multipoles.

Usually, toroidal moments are related to the radiation fields \cite{Dubovik&Tosunyan:1983,Dubovik&Tugushev:1990,Radescu&Vaman:2002,Rovenchak&Krynytskyi:2018}. We present a rigorous and consistent approach to the definitions of the three types of multipoles, namely, electric, magnetic, and toroidal multipole moments, which remains within the quasistatic approximation \cite{Larsson:2007} and differs in this regard from most works briefly analyzed above. Moreover, closed-form expressions derived in an arbitrary order for three multipole families -- electric, magnetic, and toroidal -- provide a full system for expansions of the electromagnetic fields. Despite lengthy derivations, the basic mathematical techniques applied are mostly limited to vector calculus, which ensures a low-effort tracking of the whole procedure.

The paper is organized as follows. Sec.~\ref{sec:quasi} contains details of the quasistatic approximation. Expressions for the scalar and vector potentials are given in Sec.~\ref{sec:potentials}. Multipole expansions for different parts of the electric and magnetic fields are given in Secs.~\ref{sec:multiEp}--\ref{sec:multiEv}, where magnetic and toroidal multipole moments are defined by using a notion of formally introduced analogs of the electric charge from the electric multipoles. Discussion in Sec.~\ref{sec:discussion} concludes the paper.

\section{Quasistatic electrodynamics}\label{sec:quasi}

In this Section, we will recall the definition of the quasistatic electrodynamics (see, e.g., \cite{Jackson:1999,Larsson:2007,Rosser:1997}). We will proceed from Maxwell's equations written in the Gaussian units:
\begin{subequations}\label{eqs:Maxwell1}
\begin{align}
&\div\E=4\pi\rho,\\[6pt]
&\div\B=0,\\[6pt]
&\rot\E=-\cdt\B,\\[6pt]
&\rot\B={4\pi\over c}\j+\cdt\E.
\end{align}
\end{subequations}
For convenience, we have introduced a shorthand notation for the time derivatives
\begin{align}
\cdt{X} \equiv {1\over c}{\partial X\over\partial t},\qquad
\cddtt{X} \equiv {1\over c^2}{\partial^2 X\over\partial t^2}.
\end{align}

From these equations the following second-order equations can be obtained:
\begin{subequations}\label{eq:dalms}
\begin{align}
&\square\,\E=4\pi\bnabla\rho+{4\pi\over c}\cdt\j\\[6pt]
&\square\,\B=-{4\pi\over c}\rot\j
\end{align}
\end{subequations}
where the d'Alembertian
\begin{align}
\square = \Delta - \frac{1}{c^2}\frac{\partial^2}{\partial t^2}.
\end{align}
Note that for charges and current densities the charge conservation law holds:
\begin{align}
\div\j+{\partial\rho\over\partial t}=0.
\end{align}
Eqs.~(\ref{eq:dalms}) have the structure of wave equations, which means that electric and magnetic fields at a given moment of time are defined by the behavior of charges and currents up to this moment in the past.

Let us write the electric $\E$ and magnetic $\B$ fields using the following splitting into potential (``p'' subscript), vortex (``v'' subscript), and radiative (``r'' subscript) terms:
\begin{subequations}
\begin{align}
\E&=\E_{\rm p} +\E_{\rm v}+\E_{\rm r},\\[6pt]
\B&=\B_{\rm v}+\B_{\rm r}.
\end{align}
\end{subequations}

The meaning of this splitting is as follows. Instead of two physical (objectively existing) fields we introduce five new ones. Therefore, we can impose three additional conditions for these new fields: we require that $\E_{\rm p}$, $\E_{\rm v}$ and $\B_{\rm v}$ be potential and vortex components of the electric and magnetic fields, respectively (note that $\B_{\rm p}\equiv0$ since $\div \B = 0$) and that the values of the fields at a given point and time be defined by coordinates, velocities, and accelerations of charges at the same moment of time. This means that the respective second-order equations should be not wave equations but Poissonian equations, with sources containing at most the second time derivatives of the charge density $\rho$ and the first derivative of the current density $\j$.
We will call the fields $\E_{\rm p}$, $\E_{\rm v}$, and $\B_{\rm v}$ the quasistatic fields, which coincides with e.\,g., the definition given in \cite{Larsson:2007} and corresponds the the so called Darwin model of quasistatics.

To find the equations for the quasistatic fields, we rewrite Eqs.~(\ref{eq:dalms}) as follows:
\begin{subequations}
\begin{align}
&\Delta\,\E_{\rm p}+\Delta\,\E_{\rm v}+\square\,\E_{\rm r}=
4\pi\bnabla\rho+\left({4\pi\over c}\cdt\j+\cddtt\E_{\rm p}\right)+\cddtt\E_{\rm v}\\[6pt]
&\Delta\,\B_{\rm v}+\square\,\B_{\rm r}=-{4\pi\over c}\rot\j+\cddtt\B_{\rm v}
\end{align}
\end{subequations}
and equate corresponding terms in left- and right-hand sides, which yields five new equations:
\begin{subequations}\label{eq:nabla-rho}
\begin{align}
&\Delta\E_{\rm p}=4\pi\bnabla\rho
&&\Delta\E_{\rm v}={4\pi\over c}\cdt\j+\cddtt\E_{\rm p}
&&\square\,\E_{\rm r}=\cddtt\E_{\rm v}\\[6pt]
&
&&\Delta\B_{\rm v}=-{4\pi\over c}\rot\j
&&\square\,\B_{\rm r}=\cddtt\B_{\rm v},\label{eq:DeltaBv=rotj}
\end{align}
\end{subequations}
This set clearly illustrates the motivation behind the proposed field splitting. Indeed, $\E_{\rm p}$, $\E_{\rm v}$, and $\B_{\rm v}$ satisfy the Poisson equations, which yield instantaneous solutions as described above, while $\E_{\rm r}$ and $\B_{\rm r}$ satisfy the wave equations and thus include retardation effects. The potential part $\E_{\rm p}$ is defined by the distribution of charges; there are no magnetic charges hence $\B_{\rm p}\equiv0$. The vortex parts $\E_{\rm v}$ and $\B_{\rm v}$, on the other hand, are defined by the motion of charges as reflected in the current distribution and time derivatives of $\E_{\rm p}$.

Under the condition that the quasistatic fields be ``tied'' to the charges and currents, so that zero sources should produce zero fields, one can obtain the following first-order equations for the quasistatic fields upon applying the divergence and rotor operations to the first two columns of (\ref{eq:nabla-rho}), while full Maxwell's equations result in equations for the radiative components as well:
\begin{subequations}\label{eq:quasi}
\begin{align}
&\div\E_{\rm p}=4\pi\rho
&&\div\E_{\rm v}=0
&&\div\E_{\rm r}=0\label{eq:3a}\\[6pt]
&
&&\div\B_{\rm v}=0
&&\div\B_{\rm r}=0\label{eq:3b}\\[6pt]
&\rot\E_{\rm p}=0
&&\rot\E_{\rm v}=-\cdt\B_{\rm v}
&&\rot\E_{\rm r}=-\cdt\B_{\rm r}
\label{eq:3c}\\[6pt]
&
&&\rot\B_{\rm v}={4\pi\over c}\j+\cdt\E_{\rm p}
&&\rot\B_{\rm r}=\cdt\E_{\rm v}+\cdt\E_{\rm r}\label{eq:3d}
\end{align}
\end{subequations}

The first two columns in (\ref{eq:nabla-rho}) and (\ref{eq:quasi}) are the equations for the quasistatic fields. Introducing $\E_{\rm qs} = \E_{\rm p}+\E_{\rm v}$ and $\B_{\rm qs} = \B_{\rm p}+\B_{\rm v}\equiv\B_{\rm v}$, we obtain for them almost Maxwell's equations:
\begin{subequations}\label{eqs:Maxwell-qs}
\begin{align}
&\div\E_{\rm qs}=4\pi\rho,\\[6pt]
&\div\B_{\rm qs}=0,\\[6pt]
&\rot\E_{\rm qs}=-\cdt\B_{\rm qs},\\[6pt]
&\rot\B_{\rm qs}={4\pi\over c}\j+\cdt\E_{\rm p},
\end{align}
\end{subequations}
but the non-potential part of the displacement current is neglected \cite{Larsson:2007}.

\begin{figure}[h]
\centerline{\includegraphics[scale=0.8]{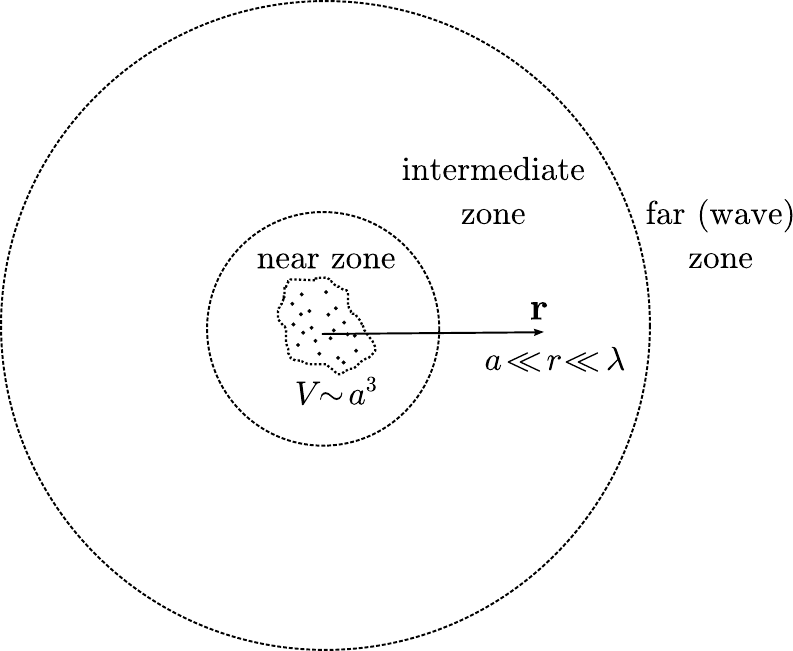}}
\caption{Typical scales. The system of charges and currents is localized to the volume $V$ with a characteristic linear size of order $a$. In the intermediate zone, the distance of the observer from the origin $r$ is much smaller than the wavelength $\lambda$ and, at the same time, much larger that the system size $a$.}\label{fig:zones}
\end{figure}

The quasistatic approximation thus corresponds to substituting Maxwell's equations with (\ref{eqs:Maxwell-qs}) and to identifying the real physical fields with $\E_{\rm qs}$ and $\B_{\rm qs}$.
The applicability of this approximation corresponds to situations when the radiative components can be neglected in comparison with the vortex ones, i.\,e., 
$\E_{\rm r}\ll\E_{\rm v}$ and $\B_{\rm r}\ll\B_{\rm v}$. Analysis of Eqs.~(\ref{eq:nabla-rho}) shows that these conditions are satisfied in the near and intermediate zones $r\ll\lambda$ (see Fig.~\ref{fig:zones}), where $\lambda=c\omega$ is a typical wavelength linked to the principal Fourier harmonic of the charge or current density. That is, the slower the change of charges and currents the larger the domain of space where the radiative fields can be neglected. In this domain, the quasistatic fields provide a high-accuracy approximation for the real fields.

Equations of the quasistatic electrodynamics are simpler than exact Maxwell's equations as the former have a consecutive structure: charges generate $\E_{\rm p}$, currents generate $\B_{\rm v}$, and alternating charges and currents do $\E_{\rm v}$. It is important to take both quasistatic parts of the electric field into account. The potential part $\E_{\rm p}$ alone, being irrotational, cannot fulfill, for instance, Faraday's law. On the other hand, radiative parts $\E_{\rm r}$ and $\B_{\rm r}$ are vortex fields (divergenceless) being thus only corrections to the respective quasistatic vortex fields.

The aim of the following sections of this work is to obtain multipole expansions for the fields $\E_{\rm p}$, $\B_{\rm v}$ and $\E_{\rm v}$, that is, to represent them as series over ${1/r}$ without imposing a condition of static charges and currents. Nevertheless, in expansions for $\E_{\rm p}$ and $\B_{\rm v}$ we see well-known series of electric and magnetic multipole moments. A nontrivial result appears in the expansion of the $\E_{\rm v}$ field, where the toroidal moments appear, which typically arise in multipole expansions of the radiation fields. It is possible, however, -- as we demonstrate -- to obtain all three types of multipole moments by expanding over ${1 /  r}$ the quasistatic fields alone.

\section{Scalar and vector potentials}\label{sec:potentials}
According to the above splitting of fields, the scalar and vector potentials are written as follows:
\begin{subequations}
\begin{align}
\phi&=\phi_{\rm p}\\[6pt]
\A&=\A_{\rm v} +\A_ {\rm r},
\end{align}
\end{subequations}
so that:
\begin{subequations}
\begin{align}
&\E_{\rm p}=-\bnabla\phi_{\rm p}
&&\E_{\rm v}=-\cdt\A_{\rm v}
&&\E_{\rm r}=-\cdt\A_{\rm r}\\[6pt]
&
&&\B_{\rm v}=\rot\A_{\rm v}
&&\B_{\rm r}=\rot\A_{\rm r}
\end{align}
\end{subequations}
For convenience, we will drop the index ``p'' of the scalar potential as $\varphi$ does not contain other terms.

Equations for the potentials are:
\begin{subequations}
\begin{align}
&\Delta\phi=-4\pi\rho
&&\Delta\A_{\rm v}=-{4\pi\over c}\j-\cdt\E_{\rm p}
&&\square\,\A_{\rm r}=\cddtt\A_{\rm v}
\label{eq:Ar_via_Av}\\[6pt]
&
&&\div\A_{\rm v}=0
&&\div\A_{\rm r}=0.
\end{align}
\end{subequations}
Note the recursive nature of these equations as well as of equations (\ref{eq:nabla-rho}). First, the potential part of the field is obtained. Being inserted into the second column, it yields the vortex part, which is sufficient for the quasistatic approximation. Finally, equations for the radiative part of fields involve the vortex part.

The solutions for the potential and vortex parts are given by:
\begin{subequations}
\begin{align}
\phi &=\int\!\!d^3\!r'{\rho'\over|\r-\r'|},\\[12pt]
\A_{\rm v} &= {1\over c}\!\int\!\! d^3\!r'
\left\{{\j'\over|\r-\r'|}+
{1\over4\pi}\frac{\partial}{\partial t}{{\E}{}'_{\rm p} \over |\r-\r'|}
\right\}
=\A_{\rm v1}+\A_{\rm v2},
\label{eq:Av=Av1+Av2}
\end{align}
\end{subequations}
where the primes at $\rho$, $\j$, and $\E_{\rm}$ denote the dependence of the integration variable $\r'$. This convention is also used in further derivations.

The first term is
\begin{align}\label{eq:Av1}
\A_{\rm v1} &= {1\over c}\!\int\!\! d^3\!r'
{\j'\over|\r-\r'|}.
\end{align}
After some transformations as given in \ref{app:A} we obtain
\begin{align*}
\A_{\rm v2}=
{1\over c}\!\int\!\!d^3\!r'\left\{-{\j'\over2|\r-\r'|}+{(\j',\r-\r')(\r-\r')\over2|\r-\r'|^3}\right\}
\end{align*}
finally arriving at the following expression for $\A_{\rm v}$:\begin{align}\label{eq:Av}
\A_{\rm v}={1\over c}\!\int\!\!d^3\!r'\left\{{\j'\over2|\r-\r'|}+{(\j',\r-\r')(\r-\r')\over2|\r-\r'|^3}\right\}.
\end{align}

\section{Multipole expansion for $\E_{\rm p}$}\label{sec:multiEp}

Recall the following relations:
\begin{align}
&{1\over|\r-\r'|}=\sum_{n=0}^\infty{(-1)^n\over n!}(\r',\bnabla)^n\,{1\over r}\\[12pt]
&\Delta\,{1\over r}=0,\qquad\Delta\,{r\over 2}={1\over r}.
\end{align}
Note that we are interested in large $r$, so there is no need in writing the delta-function term in the first Laplacian.

The potential part of the field is:
\begin{align}
&\E_{\rm p}=-\bnabla\phi.
\end{align}
The exact formula for the scalar potential
\begin{align}
\phi=\int\!\!d^3\!r'{\rho'\over|\r-\r'|}
\end{align}
yields the series:
\begin{align}
\phi=\sum_{n=0}^\infty{(-1)^n\over n!}\!\int\!\!d^3\!r'\rho'(\r',\bnabla)^n\,{1\over r}=
\sum_{n=0}^\infty{(-1)^n\over n!}
\sum_{i_1=1}^3\! \ldots\! \sum_{i_n=1}^3
Q_{i_1\ldots i_n}
\left(
\frac{\partial}{\partial x_{i_1}}\ldots \frac{\partial}{\partial x_{i_n}}
{1\over r}
\right),
\end{align}
where electric multipole moments are:
\begin{align}\label{eq:Q-electric}
Q_{i_1\dots i_n}=\int\!\! d^3\!r'\,\rho'x'_{i_1}\dots x'_{i_n}.
\end{align}
Obviously, the zero-rank tensor $Q$ is just the total electric charge and the first-rank tensor $Q_i$ is nothing but the dipole moment of the system:
\begin{align}
q=\int\!\! d^3\!r'\,\rho',\qquad
\d=\int\!\!d^3\!r'\,\r'\rho'.
\end{align}
For simplicity, we do not detrace the $Q_{i_1\dots i_n}$ tensors for $n>1$. That is, for instance, the quadrupole moment is just
\begin{align}\label{eq:Qij}
Q_{ik}=\int\!\! d^3\!r'\,\rho'x'_{i}x'_{k},
\end{align}
but not the traceless
\begin{align}\label{eq:Qij-traceless}
\bar Q_{ik}=\int\!\! d^3\!r'\,\rho'\left(x'_{i}x'_{k}-\frac{1}{3}\delta_{ik}r'^2\right),
\end{align}
where $\delta_{ij}$ is Kornecker's delta. The scalar potential $\varphi$ is independent of the multipole moment definition: one can equivalently use both the above form (\ref{eq:Q-electric}) and its traceless version due to the condition $\Delta\frac{1}{r}=0$.

\section{Multipole expansion for $\B_{\rm v}$}\label{sec:multiBv}
Since we consider a physical system, in which charges and currents are localized, in the intermediate zone (see Fig.~\ref{fig:zones}) $\j=0$ and the equations for the ``vortex'' part of the magnetic field become:
\begin{subequations}
\begin{align}
\div\B_{\rm v}&=0\\
\rot\B_{\rm v}&=-\bnabla\cdt\phi
\end{align}
\end{subequations}

Vector $\B_{\rm v}$ can be represented in the following form:
\begin{align}
\B_{\rm v}=-\bnabla\psi+\rot\cdt\C.
\end{align}
Here, the $\rot\C$ field should depend on electric moments and cause nontrivial $\rot\B$. It also should not depend on the detracing procedure of the electric multipoles.
The $\psi$ field is known as the scalar magnetic potential. Its multipole coefficients are called magnetic multipole moments (by analogy with the electric moments of $\phi$).

Requiring that $\Delta\,\C=0$, the following equations hold for $\psi$ and $\C$:
\begin{align}
\Delta\,\psi &=0,\\[6pt]
\div\C&=-\phi+{q\over r}.
\end{align}

We can choose the form of $\C$ as follows:
\begin{align}
\C&=-\sum_{n=1}^\infty{(-1)^n\over n!}\!
\int\!\!d^3\!r'\rho'\r'(\r',\bnabla)^{n-1}\,{1\over r},\\[6pt]
\rot\C&=-\sum_{n=1}^\infty{(-1)^n\over n!}\!
\int\!\!d^3\!r'\rho'[\bnabla,\r'](\r',\bnabla)^{n-1}\,{1\over r}.
\end{align}
From Eq.~(\ref{eq:DeltaBv=rotj}) we can obtain the following exact formula for $\B_{\rm v}$:
\begin{align}
\B_{\rm v}={1\over c}\!\int\!\!d^3\!r'{\rot'\j'\over|\r-\r'|}
\end{align}
For the scalar magnetic potential $\psi$ one has:
\begin{align}\label{eq:nabla-psi}
-\bnabla\psi=\B_{\rm v}-\rot\cdt\C.
\end{align}

For the $\psi$ potential we obtain (see \ref{app:B}):
\begin{align}
\psi=\sum_{n=0}^\infty{(-1)^n\over n!}\!
\int\!\!d^3\!r'
\left\{-{1\over c}{\div'[\r',\j']\over n+1}\right\}(\r',\bnabla)^n\,
{1\over r}
\end{align}
So, the magnetic scalar potential is now formally written in a form similar to the multipole expansion of the electric scalar potential, namely,
\begin{align}
\psi=\sum_{n=0}^\infty{(-1)^n\over n!}\!
\int\!\!d^3\!r'\rho'_{\rm m}(n)(\r',\bnabla)^n\,{1\over r},
\end{align}
where a certain effective $n$-dependent ``magnetic charge'' appears:
\begin{align}
\rho'_{\rm m}(n)=-{1\over c}{\div'[\r',\j']\over n+1}.
\end{align}

Consequently, magnetic multipole moments occur naturally as follows:
\begin{align}
M_{i_1\dots i_n}=\int\!\! d^3\!r'\rho'_{\rm m}(n)x'_{i_1}\dots x'_{i_n}.
\end{align}
The first three of them are
\begin{align}
M&=0,\\[6pt]
\m&={1\over 2c}\!\int\!\!d^3\!r'\,[\r',\j'],\\[6pt]
M_{ik}&={1\over 3c}\!\int\!\!d^3\!r'\,\left\{[\r',\j']_i\,x'_k+[\r',\j']_k\,x'_i\right\}
\end{align}
coinciding with the zero magnetic monopole, standard magnetic dipole moment, and magnetic quadrupole moment, respectively.


\section{Multipole expansion of $\E_{\rm v}$}\label{sec:multiEv}

Equations in the intermediate zone for the ``vortex'' part of the electric field are:
\begin{align}
&\E_{\rm v}=-\cdt\A_{\rm v}\\
&\div\A_{\rm v}=0\\
&\rot\A_{\rm v}=\B_{\rm v}=
-\bnabla\psi+\rot\cdt\C.
\end{align}

The vector potential is represented as:
\begin{eqnarray}
&&\A_{\rm v}=-\bnabla\xi+\rot\F+\Big(\cdt\C+\cdt{\C}{}'+\cdt{\C}{}''\Big).
\end{eqnarray}

The $\rot\F$ field should generate $(-\bnabla\psi)$ in $\B_{\rm v}$, therefore, it depends on the magnetic moments and it should be invariant with respect to their detracing. The third term is purely electric. The $\C$ part appears naturally, $\C'$ is required to cancel the $\div\C$ term, while $\C''$ ensures invariance with respect to the detracing of electric multipole moments. The $\bnabla\xi$ term is a ``defect'' generating a third type of multipole moments, which are called toroidal moments.

Let the following conditions hold for the constituents of the $\E_{\rm v}$ field:
\begin{align}
\Delta\,\F=0,&\\[12pt]
\div\C'=-\div\C,&\qquad
\rot\C'=0,\\[12pt]
\div\C''=0,&\qquad
\rot\C''=0.
\end{align}
Equations for $\xi$ and $\F$ become:
\begin{align}
\Delta\,\xi=0,
\end{align}
\begin{align}
\div\F=-\psi.
\end{align}

To conform with previous considerations we choose $\F,\C,\C',\C''$ in the following forms:
\begin{align}
\F&=-\sum_{n=1}^\infty{(-1)^n\over n!}\!\int\!\!d^3\!r'\rho'_{\rm m}(n)\r'(\r',\bnabla)^{n-1}\,{1\over r},
\label{eq:F}\\[6pt]
\C&=\sum_{n=0}^\infty{(-1)^n\over (n+1)!}\!
\int\!\!d^3\!r'\rho'\r'(\r',\bnabla)^n\,{1\over r},
\label{eq:C}\\[6pt]
\C'&=-\bnabla\sum_{n=0}^\infty{(-1)^n\over (n+1)!}\!\int\!\!d^3\!r'\rho'(\r',\bnabla)^{n+1}\,{r\over 2},
\label{eq:C'}\\[6pt]
\C''&=\bnabla\sum_{n=1}^\infty\alpha_n{(-1)^n\over (n+1)!}\!
\int\!\!d^3\!r'\rho'r'^2(\r',\bnabla)^{n-1}\,{1\over r},
\label{eq:C''}
\end{align}
where $\alpha_n$ are unknown coefficients. Details of their calculation as well as derivation of the $\xi$ field are given in \ref{app:C}. 

As a result we obtain:
\begin{align}
\xi=-{1\over c}\sum_{n=1}^\infty{(-1)^n\over n!}\!
\int\!\!d^3\!r'\left({2(\j',\r')\r'-(n+2)r'^2\j'\over2(2n+1)},\bnabla'\right)(\r',\bnabla)^{n-1}\,{1\over r}.
\end{align}

Shifting the summation index by unity, from $n$ to $n+1$, we finally arrive at:

\begin{itemize}
\item toroidal scalar potential:
\begin{align}
\xi=\sum_{n=0}^\infty{(-1)^n\over n!}\!
\int\!\!d^3\!r'\rho'_{\rm t}(n)(\r',\bnabla)^n\,{1\over r};
\end{align}

\item toroidal ``charge'':
\begin{align}
\rho'_{\rm t}(n)=-{1\over c}{\div'\{2(\j',\r')\r'-(n+3)r'^2\j'\}
\over 2(n+1)(2n+3)};
\end{align}

\item toroidal multipole moments:
\begin{align}
T_{i_1\dots i_n}=\int\!\! d^3\!r'\rho'_{\rm t}(n)x'_{i_1}\dots x'_{i_n}.
\end{align}

\end{itemize}

The first three toroidal moments are as follows:
\begin{align}
T&=0,\\[12pt]
\T&={1\over 10c}\!\int\!\!d^3\!r'\{(\j',\r')\r'-2r'^2\j'\},\\[12pt]
T_{ik}&={1\over 42c}\!
\int\!\!d^3\!r'\{4(\j',\r')x'_ix'_k-5r'^2(j'_ix'_k+j'_k x'_i)\}.
\end{align}
The second expression, for vector $\T$, is nothing but the so called toroidicity. Note that, up to integer multipliers, the inverse factor of ${42}$ coincides, e.\,g., with definitions in \cite{Alaee_etal:2018,Porsev:1994,Vrejoiu&Nicmorus:2004,Talebi_etal:2018} or differs by $\frac{3}{2}$ from ${28}$ as given in \cite{Dubovik&Cheshkov:1974}. As a relief after all the conducted derivations recall that ``42'' in the third moment is also the Answer to the Ultimate Question of Life, the Universe, and Everything \cite{Adams:1979}.

\section{Discussion}\label{sec:discussion}

We have proposed a consistent approach to the definition of multipole moments occurring in expansions of the electromagnetic field generated by time-dependent distributions of charges and currents. Upon splitting the fields into potential, vortex, and radiative terms we have managed to derive general expressions for three types of moments: electric, magnetic, and toroidal. This became possible by means of formally introduced magnetic and toroidal analogs of the electric charge with subsequent application of techniques similar to the multipole expansion of the electrostatic field in terms of the scalar potential.

The proposed scheme is summarized in Table~\ref{tab:summary}. Electric, magnetic, and toroidal multipole moments appear in expansions of the scalar fields $\varphi$, $\psi$, and $\xi$ under gradients in $\E_{\rm p}$, $\B_{\rm v}$, and $\A_{\rm v}$, respectively. The transitions from $\A_{\rm v}$ back to $\E_{\rm p}$ are made via consecutive applications of rotors. Vectors $\rot \F$ and $\C, \C',\C''$ are expressed via multipole moments in the corresponding columns, i.e., magnetic and electric ones, respectively.

\vspace*{-1ex}
\begin{table}[H]
\caption{General scheme for obtaining three types of multipoles.}
\label{tab:summary}

\vspace*{-3pt}
\begin{center}
\begin{tabular}{rccccccc}
\hline\\[-12pt]
& & & Toroidal && Magnetic && Electric \\
& & &moments&&moments&&moments\\
\hline\\
& $\displaystyle \cdt\E_{\rm p}$ &=&
& & & &
$\displaystyle -\bnabla\cdt\varphi$\\[8pt]
$\rot$: & $\Uparrow$\\
& $\B_{\rm v}$ &=& & & $-\bnabla\psi$ &
$+$
& $\displaystyle \rot\cdt\C$\\[8pt]
$\rot$: & $\Uparrow$\\
& $\A_{\rm v}$ &=& $-\bnabla\xi$ & $+$ & $\rot\F$ & $+$ &
$\displaystyle \Big(\cdt\C+\cdt{\C}{}'+\cdt{\C}{}''\Big)$\\[8pt]
\hline
\end{tabular}
\end{center}
\end{table}

\vspace*{-2ex}
Explicit results for first three orders in the multipole expansions of fields are presented in \ref{app:D}. Given the distribution of charges and currents, those formulas can be directly applied for calculations.

Some general observations regarding our approach can be formulated as follows:

\begin{itemize}
\item The splitting of $\B_{\rm v}$ into magnetic ($-\bnabla\psi$) and electric ($\rot\cdt\C$) parts
is unique up to the choice of the radius-vector origin.

\item The same holds true for the splitting of $\A_{\rm v}$ (and hence $\E_{\rm v}$) into toroidal ($-\bnabla\xi$), magnetic ($\rot\F$), and electric ($\cdt\C+\cdt{\C}{}'+\cdt{\C}{}''$) parts.

\item All the fields in the right-hand sides in Table~\ref{tab:summary} ($-\bnabla\phi$,  $-\bnabla\psi$,  $-\bnabla\xi$, $\rot\F$, $\rot\C$, $\C+\C'+\C''$) do not depend on the types of moments, traceless or not.

\item The series for electric, magnetic, and toroidal moments are generated by the large-$r$ expansions of quasistatic fields $\E_{\rm p}$, $\B_{\rm v}$, and  $\E_{\rm v}$ only.

\item The magnetic and toroidal moments together with time-derivatives of electric moments form a linearly independent and full system, i.e., they form a basis for current moments (while electric moments correspond to charge). So, there is no need to introduce other types of multipoles -- they will appear as linear combinations of the three defined moments.

\item The coefficients in expansions of the radiative fields $\A_{\rm r}$, $\B_{\rm r}$ and $\E_{\rm r}$ at large $r$ are unambiguously expressed via the found moments, their traces and time derivatives (the same holds true for the radiation intensity).

\end{itemize}

In summary, we expect that the proposed approach will take its proper place in the theory of multipole expansion of time-dependent charge and current distributions, even though this is a problem in classical electrodynamics with a long history and a number of definitions known so far.

\section*{Acknowledgments}\vspace*{-1ex}
We are grateful to Dr. Mykola Stetsko for reading the manuscript.
Comments from the anonymous Referee are highly acknowledged.

\appendix
\renewcommand{\theequation}{\thesection.\arabic{equation}}
\setcounter{equation}{0}
\section{Deriving $\A_{\rm v2}$}\label{app:A}

The second term $\A_{\rm v2}$ of the vector potential (\ref{eq:Av=Av1+Av2}) can be written via currents after some transformations. First of all, it is straightforward to show that
\begin{align}
\A_{\rm v2}=
{1\over c}\!\int\!\!d^3\!r'\,
{1\over4\pi}{\partial\E'_{\rm p}\over\partial t}{1\over|\r-\r'|}
=-{1\over c}\!\int\!\!d^3\!r'\,
{1\over4\pi}
{\partial\E'_{\rm p}\over\partial t}\div'{(\r-\r')\over2|\r-\r'|}.
\end{align}
The $\div'$ operator acts on the $\r'$ variable. Upon applying the following vector identity,
\begin{align}
(\bnabla\a)\b+(\bnabla\b)\a-\bnabla(\a\b) =
[\rot\a,\b]+[\rot\b,\a]+\a\div\b+\b\div\a,
\end{align}
one can integrate the expression for $\A_{\rm v2}$ by parts. Let $\a = \E_{\rm p}$ and $\b = \frac{(\r-\r')}{2|\r-\r'|}$, so that $\rot\a=\rot\b=0$. The terms with $\bnabla$ become surface integrals vanishing -- here and throughout further derivations -- at large distances. So, making some manipulations,
\begin{align*}
&\A_{\rm v2}={1\over c}\!\int\!\!d^3\!r'\,{1\over4\pi}{\partial\div'\E'_p\over\partial t}{\r-\r'\over2|\r-\r'|}={1\over c}\!\int\!\!d^3\!r'\,{\partial\rho'\over\partial t}{\r-\r'\over2|\r-\r'|}=\\[6pt]
&={1\over c}\!\int\!\!d^3\!r'\,(-\div'\j'){\r-\r'\over2|\r-\r'|}=
{1\over c}\!\int\!\!d^3\!r'\,(\j',\bnabla'){\r-\r'\over2|\r-\r'|}=\\[6pt]
&={1\over c}\!\int\!\!d^3\!r'\left\{-{\j'\over2|\r-\r'|}+{(\j',\r-\r')(\r-\r')\over2|\r-\r'|^3}\right\}
\end{align*}

\setcounter{equation}{0}
\section{Deriving $\psi$}\label{app:B}
The terms in the right-hand side of (\ref{eq:nabla-psi}) are expressed as series:
\begin{align}
\B_{\rm v}&={1\over c}\sum_{n=0}^\infty{(-1)^n\over n!}\!\int\!\!d^3\!r'\rot'\j'(\r',\bnabla)^n\,{1\over r}\nonumber\\[6pt]
&={1\over c}\sum_{n=1}^\infty{(-1)^n\over n!}\!\int\!\!d^3\!r'[\j',\bnabla'](\r',\bnabla)^n\,{1\over r}\\[12pt]
\rot\cdt\C
&=-{1\over c}\sum_{n=1}^\infty{(-1)^n\over n!}\!\int\!\!d^3\!r'{\partial\rho'\over\partial t}[\bnabla,\r'](\r',\bnabla)^{n-1}\,{1\over r}\nonumber\\
&=-{1\over c}\sum_{n=1}^\infty{(-1)^n\over n!}\!\int\!\!d^3\!r'(\j',\bnabla')[\bnabla,\r'](\r',\bnabla)^{n-1}\,{1\over r}
\end{align}

The series for $-\bnabla\psi$ becomes:
\begin{align}\label{eq:nabla_psi=}
-\bnabla\psi={1\over c}\sum_{n=1}^\infty{(-1)^n\over n!}\!\int\!\!d^3\!r'
\left\{[\j',\bnabla'](\r',\bnabla)+(\j',\bnabla')[\bnabla,\r']\right\}
(\r',\bnabla)^{n-1}\,{1\over r}.
\end{align}
We will normalize the operator in curly braces by moving the primed nablas to the rightmost positions:
\begin{align}
\{\dots\}&=
[\j',\bnabla]+(\bnabla,\r')[\j',\bnabla']+[\bnabla,\j']+
[\bnabla,\r'](\j',\bnabla')\nonumber\\[6pt]
&=(\bnabla,\r')[\j',\bnabla']+[\bnabla,\r'](\j',\bnabla').
\end{align}
It is now convenient to use the vector identity
\begin{align}\label{eq:vector_identity}
(\a\b)[\cvec\d]=(\a\cvec)[\b\d]+(\a\d)[\cvec\b]+\a([\b\cvec]\d)
\end{align}
following from two different ways of writing the expression $[\b[\a[\cvec\d]]]$, where $\a$ corresponds to $\bnabla$ in the first case and to $\j'$ in the second one:
\begin{align}
\{\dots\}
&=(\bnabla,\j')[\r',\bnabla']+[\j',\r'](\bnabla,\bnabla')+ \bnabla([\r',\j'],\bnabla')\nonumber\\[6pt]
&+(\j',\r')[\bnabla,\bnabla']-
(\bnabla,\j')[\r',\bnabla']+\j'([\bnabla,\r'],\bnabla')\nonumber\\[6pt]
&=[\j',\r'](\bnabla,\bnabla')+\bnabla([\r',\j'],\bnabla')+
(\j',\r')[\bnabla,\bnabla']-\j'(\r',[\bnabla,\bnabla'])
\end{align}

One can easily check that $(\bnabla,\bnabla')$ and $[\bnabla,\bnabla']$ in Eq.~(\ref{eq:nabla_psi=}) yield zero contribution, thus only the second term remains:
\begin{align}
-\bnabla\psi&=
{1\over c}\sum_{n=1}^\infty{(-1)^n\over n!}\!
\int\!\!d^3\!r' \bnabla([\r',\j'],\bnabla')(\r',\bnabla)^{n-1}\,{1\over r}
\nonumber\\[6pt]
&=
{1\over c}\bnabla\sum_{n=1}^\infty{(-1)^n\over n!}\!
\int\!\!d^3\!r' (-\div'[\r',\j'])(\r',\bnabla)^{n-1}\,{1\over r}
\end{align}
Shifting the summation index by unity, from $n$ to $(n+1)$, we finally get:
\begin{align}
\psi=\sum_{n=0}^\infty{(-1)^n\over n!}\!
\int\!\!d^3\!r'
\left\{-{1\over c}{\div'[\r',\j']\over n+1}\right\}(\r',\bnabla)^n\,
{1\over r}
\end{align}

\setcounter{equation}{0}
\section{Deriving $\xi$}\label{app:C}
In Eq.~(\ref{eq:C''}), the only free parameters are $\alpha_n$. They should be fixed from the condition that the sum of fields $\C+\C'+\C''$ is independent of the detracing procedure for the electric multipole moments. So, inserting into the expressions for  $\C,\C',\C''$ at given $n\ge1$ instead of
\begin{align*}
\rho'x'_{i_1}\dots x'_{i_{n+1}}
\end{align*}
``detracers'' given by
\begin{align*}
\sum_{\{j_1,j_2\}\subset\{i_1,\dots,i_{n+1}\}}\rho'r'^2\delta_{j_1j_2}x'_{k_1}\dots x'_{k_{n-1}}
\end{align*}
should yield zero. In the above expression, the summation runs over all the pairs of indices $\{j_1,j_2\}$ within given $\{i_1,\dots,i_{n+1}\}$, while $\{k_1,\dots,k_{n-1}\}$ are the remaining indices; the total number of summands is $n(n+1)/2$.

We thus have an equation for $\alpha_n$ as follows:
\begin{align}
0=\bnabla\sum_{n=1}^\infty{(-1)^n\over (n+1)!}\!
\int\!\!d^3\!r'\rho'r'^2
\left\{n-{n(n+1)\over2}+\alpha_n[3+2(n-1)]\right\}
(\r',\bnabla)^{n-1}\,{1\over r}
\end{align}
and
\begin{align}
\alpha_n={(n-1)n\over 2(2n+1)}.
\end{align}

The series for $\C'$ can be written in the following integral form:
\begin{align}
\C'=\int\!\!d^3\!r'\rho'\sum_{n=0}^\infty{(-1)^{n+1}\over (n+1)!}
(\r',\bnabla)^{n+1}\,{\r\over2r}=
\int\!\!d^3\!r'\rho'{\r-\r'\over2|\r-\r'|}-q{\r\over2r}.
\end{align}
Therefore,
\begin{align}
\cdt\C{}'=
\!\int\!\!d^3\!r'\,
\cdt\rho{}'{\r-\r'\over2|\r-\r'|}=\A_{\rm v2}
\end{align}

The equation for $\xi$ becomes:
\begin{align}
-\bnabla\xi=\A_{\rm v1}-\rot\F-
\Big(\cdt\C+\cdt\C{}''\Big).
\end{align}

Taking into consideration series (\ref{eq:Av1}), (\ref{eq:F})--(\ref{eq:C''}) for $\A_{\rm v1}$, $\rot\F$, $\cdt\C$, and $\cdt\C{}''$, expressed via currents:
\begin{align}
\A_{\rm v1}&={1\over c}\sum_{n=0}^\infty{(-1)^n\over n!}\!\int\!\!d^3\!r'\j'(\r',\bnabla)^n\,{1\over r}=\nonumber\\[6pt]
&={1\over c}\sum_{n=1}^\infty{(-1)^n\over n!}\!
\int\!\!d^3\!r'\j'(\r',\bnabla)^n\,{1\over r}+
{1\over cr}\!\int\!\!d^3\!r'\j',\\[12pt]
\rot\F&=-{1\over c}\sum_{n=1}^\infty{(-1)^n\over (n+1)!}\!\int\!\!d^3\!r'([\r',\j'],\bnabla'))[\bnabla,\r'](\r',\bnabla)^{n-1}\,
{1\over r},\\[12pt]
\cdt\C
&=
{1\over c}\sum_{n=0}^\infty{(-1)^n\over (n+1)!}\!\int\!\!d^3\!r'
(\j',\bnabla')\r'(\r',\bnabla)^n\,{1\over r}\nonumber\\[6pt]
&={1\over c}\sum_{n=1}^\infty{(-1)^n\over (n+1)!}\!\int\!\!d^3\!r'
(\j',\bnabla')\r'(\r',\bnabla)^n\,{1\over r}+{1\over cr}\!\int\!\!d^3\!r'\j',\\[12pt]
\cdt\C{}''
&=
\bnabla{1\over c}\sum_{n=1}^\infty{(-1)^n\over (n+1)!}\!
\int\!\!d^3\!r'
{(n-1)n\over 2(2n+1)}(\j',\bnabla')r'^2(\r',\bnabla)^{n-1}\,{1\over r},
\end{align}
we obtain the series for $(-\bnabla\xi)$:
\begin{align}
-\bnabla\xi={1\over c}\sum_{n=1}^\infty{(-1)^n\over(n+1)!}\!
\int\!\!d^3\!r'
&\left\{(n+1)\j'(\r',\bnabla)+
([\r',\j'],\bnabla')[\bnabla,\r']-
(\j',\bnabla')\r'(\r',\bnabla)
\right\}\nonumber\\
&\times
(\r',\bnabla)^{n-1}{1\over r}-\cdt\C{}''
\end{align}

The $\C''$ vector is a gradient itself, so we will work with it later. The first term in curly braces together with the external factor of $(\r',\bnabla)^{n-1}$ might be written as follows in order to get rid of $n$:
\begin{align}
(n+1)\j'(\r',\bnabla)^n=n\j'(\r',\bnabla)^n+\j'(\r',\bnabla)^n=
\j'(\r,'\bnabla')(\r',\bnabla)^n+\j'(\r',\bnabla)^n
\end{align}
The operator in curly braces becomes:
\begin{align}
\{\dots\}=\j'(\r,'\bnabla')(\r',\bnabla)+\j'(\r',\bnabla)+
([\r',\j'],\bnabla')[\bnabla,\r']-(\j',\bnabla')\r'(\r',\bnabla).
\end{align}
As before, it can be normalized by moving the primed nablas to the rightmost positions:
\begin{align*}
&\{\dots\}=2\j'(\r',\bnabla)+\j'(\r',\bnabla)(\r,'\bnabla')
+[\bnabla,[\r',\j']]+[\bnabla,\r']([\r',\j'],\bnabla')-\\[6pt]
&-\j'(\r',\bnabla)-\r'(\j',\bnabla)-\r'(\r',\bnabla)(\j',\bnabla')=\\[6pt]
&=\j'(\r',\bnabla)(\r',\bnabla')+[\bnabla,\r']([\r',\j'],\bnabla')-\r'(\r',\bnabla)(\j',\bnabla').
\end{align*}
Upon applying identity (\ref{eq:vector_identity}) to the second term choosing $\a$ as $[\r',\j']$ we obtain:
\begin{align}
[\bnabla,\r']([\r',\j'],\bnabla')=
([\r',\j'],\r')[\bnabla,\bnabla']-([\r',\j'],\bnabla)[\r',\bnabla']-
[\r',\j'](\r',[\bnabla,\bnabla']).
\end{align}

Since the operator $[\bnabla,\bnabla']$ acts trivially, as we have seen before, we can again transform only the second term choosing $\a$ equal to $\bnabla$:
\begin{align}
-([\r',\j'],\bnabla)[\r',\bnabla']=-[\r',[\r',\j']](\bnabla,\bnabla')-
(\r',\bnabla)[[\r',\j'],\bnabla']-\bnabla([\r',\j'],[\r',\bnabla']).
\end{align}

The action of $(\bnabla,\bnabla')$ is also trivial, so only the second and third terms enter into $\{\dots\}$:
\begin{align}
\{\dots\}&=\j'(\r',\bnabla)(\r',\bnabla'))-(\r',\bnabla)[[\r',\j'],\bnabla']- \bnabla([\r',\j'],[\r',\bnabla'])-\r'(\r',\bnabla)(\j',\bnabla')
\nonumber\\[6pt]
&=-\bnabla([\r',\j'],[\r',\bnabla'])=\bnabla([\r',[\r',\j']],\bnabla').
\end{align}
Only non-trivial contributions are preserved under transition to the last equality.

Let us now write the series for $(-\bnabla\xi)$ taking into account $\C''$:
\begin{align}
-\bnabla\xi=\bnabla{1\over c}\sum_{n=1}^\infty
{(-1)^n\over (n+1)!}\!
\int\!\!d^3\!r'\left\{([\r',[\r',\j']],\bnabla')-
{(n-1)n\over 2(2n+1)}(\j',\bnabla')r'^2\right\}(\r',\bnabla)^{n-1}\,{1\over r}.
\end{align}
Normalization of the second term gives:
\begin{align}
(\j',\bnabla')r'^2=r'^2(\j',\bnabla')+2(\j',\r').
\end{align}
The last term together with $(\r',\bnabla)^{n-1}$ can be transformed as follows (note that here $n>1$):
\begin{align}
2(\j',\r')(\r',\bnabla)^{n-1}={2\over n-1}(\j',\r')(\r',\bnabla')(\r',\bnabla)^{n-1}.
\end{align}

The expression for $\xi$ is now:
\begin{align}
\xi=-{1\over c}\sum_{n=1}^\infty{(-1)^n\over (n+1)!}\!
&\int\!\!d^3\!r'
 \bigg\{([\r',[\r',\j']],\bnabla')- \nonumber \\
& -{(n-1)n\over 2(2n+1)}\left(r'^2(\j',\bnabla')+
{2\over n-1}(\j',\r')(\r',\bnabla')\right)\bigg\}(\r',\bnabla)^{n-1}\,{1\over r}.
\end{align}
The curly braces can be written in the form of $(\dots,\bnabla')$, where the left argument is:
\begin{align}
\dots&={2(2n+1)[\r',[\r',\j']]-n(n-1)r'^2\j'-2n(\j',\r')\r'\over2(2n+1)}
\nonumber\\
&=
{2(n+1)(\j',\r')\r'-(n+1)(n+2)r'^2\j'\over2(2n+1)}.
\end{align}

\setcounter{equation}{0}
\section{First orders in multipole expansions}\label{app:D}
Electric and magnetic fields corresponding to the quasistatic approximations are given by:
\begin{subequations}\label{eq:EpEvBv}
\begin{align}
\E_{\rm p}&=-\bnabla\phi;\\[6pt]
\E_{\rm v}&=\phantom{-}\bnabla\cdt\xi-\rot\cdt\F-
\Big( \cddtt\C + \cddtt\C{}' + \cddtt\C{}'' \Big);\\
\B_{\rm v}&=-\bnabla\psi+\rot\cdt\C.
\end{align}
\end{subequations}
The first three terms in the multipole expansions of these fields are explicitly written in Table~\ref{tab:appD}.

\begin{table}[H]
\caption{Multipole expansions of terms contained in Eqs.~(\ref{eq:EpEvBv})}
\label{tab:appD}

\begin{center}
\begin{tabular}{ccccc}
\hline
\multicolumn{2}{c}{
\backslashbox{Field}{$r^{-n}\vphantom{\int^M}$}
} &
$n=1$ & $n=2$ & $n=3$ \\
\hline\\
$\E_{\rm p}$: &
$-\nabla_i\varphi$ & ------ & $-q \nabla_i\left(\frac{1}{r}\right)$ &
$d_j \nabla_j\nabla_i\left(\frac{1}{r}\right)$
\\
\\
\hline
\hline
\\
\multirow{2}{*}[-1ex]{$\B_{\rm v}$:} &
$-\nabla_i\psi$ & ------ & ------ &
$m_j \nabla_j\nabla_i\left(\frac{1}{r}\right)$
\\[12pt]
& $({\rot\cdt\C})_i$ & ------ &
$\varepsilon_{ijk}\cdt d_k\nabla_j\left(\frac{1}{r}\right)$ &
$-\frac{1}{2!}\varepsilon_{ijk}\cdt Q_{kl}\nabla_l\nabla_j\left(\frac{1}{r}\right)$
\\
\\
\hline
\hline
\\
\multirow{5}{*}[-5ex]{$\E_{\rm v}$:} &
$\nabla_i\cdt\xi$ & ------ & ------ &
$-\cdt t_j \nabla_j\nabla_i\left(\frac{1}{r}\right)$
\\[12pt]
& $-({\rot\cdt\F})_i$ & ------ &
$-\varepsilon_{ijk}\cdt m_k\nabla_j\left(\frac{1}{r}\right)$ &
$\frac{1}{2!}\varepsilon_{ijk}\cdt M_{kl}\nabla_l\nabla_j\left(\frac{1}{r}\right)$
\\[12pt]
& $-\cddtt C_i$ & $-\cddtt d_i \, \frac{1}{r}$ &
$\frac{1}{2!}\cddtt Q_{ij}\nabla_j\left(\frac{1}{r}\right)$ &
$-\frac{1}{3!}\cddtt O_{ijk}\nabla_j\nabla_k\left(\frac{1}{r}\right)$
\\[12pt]
& $-\cddtt{C'_i}$ & $\cddtt d_j \nabla_j\nabla_i \left(\frac{r}{2}\right)$ &
$-\frac{1}{2!}\cddtt Q_{jk}\nabla_j\nabla_k\nabla_i \left(\frac{r}{2}\right)$ &
$\frac{1}{3!}\cddtt O_{jkl}\nabla_j\nabla_k\nabla_l\nabla_i\left(\frac{r}{2}\right)$
\\[12pt]
& $-\cddtt{\,C''_i}$ & ------ & ------ &
$-\frac{1}{5}\cdot\frac{1}{3!}\cddtt O_{jjk}\nabla_k\nabla_i\left(\frac{1}{r}\right)$
\\[8pt]
\hline
\end{tabular}
\end{center}
\end{table}

\vspace*{-1ex}
In this Table, $q$ is the total electric charge of the system, $d_j$ is the $j$th component of the electric dipole moment, $Q_{ij}$ and $O_{ijk}$ are the components of the electric quadrupole and octupole moments, respectively;
$m_j$ and $M_{ij}$ are the components of the magnetic dipole and quadrupole moments, respectively; $t_j$ is the toroidal dipole moment. $\nabla_i$ corresponds to the partial derivative $\partial/\partial x_i$; the notation $\varepsilon_{ijk}$ is used for the Levi-Civita symbol, and the summation over the repeating indices is implied.


\end{document}